\documentclass[conference]{IEEEtran}
\IEEEoverridecommandlockouts
\usepackage{cite}
\usepackage{amsmath,amssymb,amsfonts}
\usepackage{algorithmic}
\usepackage{graphicx}
\usepackage{pifont}
\newcommand{\cmark}{\ding{51}}%
\newcommand{\xmark}{\ding{55}}
\usepackage{textcomp}
\usepackage{xcolor}
\usepackage{listings}
\usepackage{lipsum}
\usepackage{verbatimbox}
\usepackage[euler]{textgreek}
\usepackage{gensymb}
\usepackage{mathrsfs}

\newcommand{\R}{\mathbb{R}}
\usepackage{mathtools}

\DeclarePairedDelimiter\floor{\lfloor}{\rfloor}
\def\BibTeX{{\rm B\kern-.05em{\sc i\kern-.025em b}\kern-.08em
    T\kern-.1667em\lower.7ex\hbox{E}\kern-.125emX}}
\usepackage{hyperref}
\hypersetup{
    colorlinks=true,
    linkcolor=magenta,
    filecolor=blue,      
    urlcolor=cyan,
    citecolor=green,
} 

\begin{document}
\title{A Robust CNN Framework with Dual Feedback Feature Accumulation for Detecting Pneumonia Opacity from Chest X-ray Images}

\author{
\IEEEauthorblockN{\textcolor{black}{Md. Jahin Alam\textsuperscript{1}, Shams Nafisa Ali\textsuperscript{2} and Md. Zubair Hasan\textsuperscript{3}}}
\IEEEauthorblockA{\text{\textcolor{black}{\textsuperscript{1}Department of Electrical and Electronic Engineering, Bangladesh University of Engineering and Technology}} \\
\text{\textcolor{black}{\textsuperscript{2}Department of Biomedical Engineering, Bangladesh University of Engineering and Technology}} \\
\text{\textcolor{black}{\textsuperscript{3}Department of Computer Science and Engineering, Military Institute of Science and Technology}} \\
\text{\textcolor{black}{ Dhaka, Bangladesh.}} \\
\textcolor{black}{\textsuperscript{1}jahinalam.eee.buet@gmail.com,
\textsuperscript{2}snafisa.bme.buet@gmail.com,
\textsuperscript{3}mdzh1997@gmail.com}}}

\maketitle

\begin{abstract}
Pneumonia is one of the most acute respiratory diseases having remarkably high prevalence and mortality rate. Chest X-ray (CXR) has been widely utilized for the diagnosis of this disease owing to its availability, diagnostic speed and accuracy. However, even for an expert radiologist, it is quite challenging to readily determine pneumonia opacity by examining CXRs. Therefore, this study has been structured to automate the pneumonia detection process by introducing a robust deep learning framework. The proposed network comprises of Process Convolution (Pro\_Conv) blocks for feature accumulation inside Dual Feedback (DF) blocks to propagate the feature maps towards a viable detection. Experimental analysis showcase: (1) the proposed network proficiently distinguishes between normal and pneumonia opacity containing CXRs with the mean accuracy, sensitivity and specificity of 97.78\%, 98.84\% and 95.04\%, respectively; (2) the network is constructed with significantly low parameters than the traditional ImageNets to reduce memory consumption for deployment in memory constrained mobile platforms; (3) the trade-off between accuracy and number of parameters of the model outperforms the considered classical networks by a remarkable margin; and (4) the false-negatives are lower than the false-positives (both of which are low in count) which prove the model's low-fatality prediction. Hence, the proposed network can be deployed for a rapid screening of pneumonia and can act as a great assistive tool for the radiologists in the diagnosis process.
\end{abstract}


\section{Introduction}
Pneumonia is a highly prevalent acute respiratory disease. It is caused by bacterial, viral, or fungal infection of lungs~\cite{bates1992microbial}. Despite the availability of advanced antibiotics, the relatively high incidence rate and mortality rate of pneumonia has invoked major public health concern around the globe~\cite{info}. For tackling this deadly disease, early diagnosis is apparently a dire need. Owing to the ease of access and enhanced depiction of discriminative features, chest X-ray (CXR) has been a widely popular diagnostic method for early screening of pneumonia~\cite{jaiswal2019}. Nevertheless, the visual differences in CXR are very subtle for various thoracic pathologies and it is quite difficult to identify an area of increased opacity from the projection film. This subsequently makes the manual screening a tedious and time-consuming work even for the expert radiologists. In this scenario, artificial intelligence (AI)-empowered computer-aided diagnostic systems come into the frame for analyzing and interpreting CXRs with substantial accuracy and efficiency. 
\par
Over the past few decades, research endeavors have been made to design automated algorithms based thoracic disease diagnosis using CXR~\cite{jaiswal2019, chapman2001comparison, rajpurkar2017chexnet, sharma2020feature, chandra2020pneumonia}. However, statistical analysis and machine learning based methods mostly require manually crafted complex features ~\cite{qin2018computer}. However, deep learning based methods are conspicuously more generalized and thereby, able to mitigate the concerns of ML-based methods~\cite{li2020accuracy}. Nevertheless, against the magnificent performance of the deep works, their resource intensive training acts as a trade-off and ultimately make them unfeasible for applying in point-of-care remote setup and mobile platforms. To resolve this issue, a large fraction of contemporary medical diagnosis algorithms are being geared towards efficient approaches namely weight quantization~\cite{xu2018quantization, acharya2020deep}, low precision~\cite{cherupally2019ecg} and lightweight networks~\cite{shuvo2020, xiao2020heart}. Although numerous investigations on CXRs have been implemented to detect pneumonia opacity, the existing works have scarcely explored~\cite{sharma2020feature} the efficient approaches considering the trade-off between accuracy and number of parameters. Therefore, the central idea behind this work involves constructing an efficient CNN architecture that is capable of yielding superior accuracy despite having significantly low parameters than the traditional ImageNets i.e., VGG19 \cite{simonyan2014very}, ResNet50 \cite{he2016deep} etc. 
\par 

\begin{figure*}[!h]
  \includegraphics[width=\textwidth]{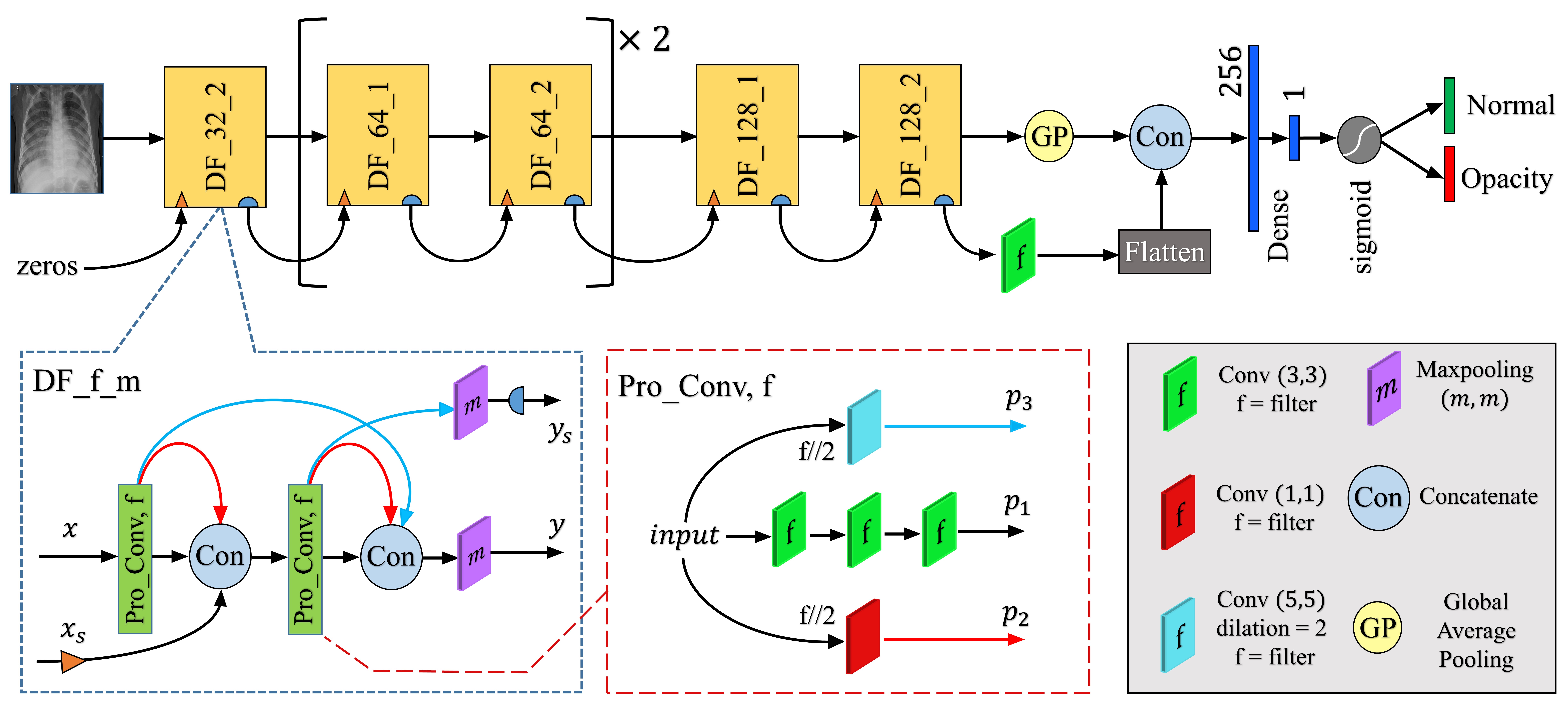}
  \caption{\textbf{Overview of the proposed scheme: The first DF block (Dual Feedback) takes in the image and zeros as input and generates two outputs with f=32 (process conv filter)  and m=2 (maxpool kernel). The rest of the DF blocks takes in pre-defined f and m values operating in the shown manner to estimate the opacity of the lung.}}
  \label{f1}
\end{figure*} 

In this paper, a robust deep convolutional neural network (CNN) architecture is proposed with the incorporation of dual feedback feature accumulation blocks (DFs) for the identification of the subtle variations amongst images with or without opacity. The proposed scheme is primarily constructed with a forward path containing simple convolutional operations to learn the significant opacity characteristics between images. Secondly, connections with smaller kernels are concatenated to assist the main forward path to preserve gradients during back-propagation as well as details of the features. Furthermore, higher dimensional kernels with dilation are introduced with bigger skips to provide spatial correlation among features which are very important for classifying images which depend on small variations between pixels. This path containing the higher dimensional kernels propagates alongside the main path both passing through the DF blocks in every single scale; thereby providing dual feedback. These DF blocks not only provide extensive feature values but also use skip-connections with multi-scale kernels for preserving detail and extracting additional receptive features.

\section{Methodology}
In this section, the problem pipeline has been formulated with an elaborate explanation of each structural part of the proposed model for detecting pneumonia opacity from chest X-ray images.

\subsection{Process Convolution Block}
Referred to as 'Pro\_Conv' in Fig. \textcolor{red}{1}, this block takes in an introduced $input\in \R^{H\times W \times C}$ and creates 3 different outputs \(p_1,\, p_2,\, p_3\)$\in \R^{H\times W \times C}$ which are formulated below:

\begin{align}
    p_1 &= \Phi_{3\_{d1}}(\Phi_{3\_d1}(\Phi_{3\_d1}(input)))
\end{align}
\begin{align}
    p_2 &= \Phi_{1\_d1}(input)
\end{align}
\begin{align}
    p_3 &= \Phi_{5\_d2}(input)
\end{align}

Here, for general use, \(\Phi_{k\_{d'N'}}\) indicates convolutional operation with \((k\times k)\) kernel and a dilation rate of `N'. The filter value is set half for \(p_2\) and \(p_3\) compared to \(p_1\). \(p_1\) provides extensive feature values for the given input whereas \(p_2\) and \(p_3\) introduce smaller and enhanced receptive field values, respectively. This block is used in the next scheme.

\subsection{Dual Feedback Block}
This block takes two inputs \(x,\, x_s\)$\in \R^{H\times W \times C}$ and outputs two feature maps \(y,\, y_s\)$\in \R^{H\times W \times C}$; hence named `Dual Feedback' as showcased in Fig.~\ref{f1} with the abbreviation `DF\_f\_m'. Here, f and m respectively indicates filter count in Pro\_Conv and maxpool kernel size before getting \(y\) and \(y_s\). Two individual Pro\_Conv blocks are utilized inside this particular scheme. The entire procedure can be written as follows:

\begin{align}
    p_{1\_1},\; p_{2\_1},\; p_{3\_1}  = Pro\_Conv_1(x)\Big|_f \\
    x_{c1} = Concat(x_s,\; p_{1\_1}, \;p_{2\_1})\\
    p_{1\_2},\; p_{2\_2},\; p_{3\_2}= Pro\_Conv_2(x_{c1})\Big|_f\\
    x_{c2} = Concat(p_{3\_1},\; p_{1\_2}, \;p_{2\_2})\\
    y = MaxPool(x_{c2})\Big|_{(m\times m)}\\
    y_s = MaxPool(p_{3\_2})\Big|_{(m\times m)}
\end{align}

Here, \(p_{i\_{j}}\) indicates the \(i\)-th output of the \(j\)-th process convolution block that arrives in sequence. Moreover, \(x_{c1}\) and \(x_{c2}\) are the feature maps obtained just after concatenating the necessary layers depicted in equation (5) and (7) respectively. 

\subsection{Proposed Model Sequence}
The model structure which is implemented for the scheme, is shown in the upper section of Fig.~\ref{f1}. A series of seven DF blocks are introduced with increasing filter values (f) and changing m value to reduce the spatial dimension as well as increase channels. At the final DF block, the \(y\) is pooled but the \(y_s\) is convolved and flattened before concatenating with each other. A dense layer is added just before the prediction layer to bring the proposed framework to completion.      

\section{Experiments}
\subsection{Experimental Setup}
\subsubsection{Dataset}
For this experiment, Chest X-ray images open sourced from Kaggle were trained on during routine check ups in Guangzhou Women and Children’s Medical Center, Guangzhou~\cite{d1}. The dataset is originally divided into train (Normal-1082, Opacity-3110) and validation (Normal-267, Opacity-773) sets. The image shapes vary between \((x=127, y=384,\; channel=3)\) to \((x=2663, y=2373,\; channel= 3)\) with means and standard deviations of (\( \bar{x}=961.056, \bar{y}= 1311.40\)) and (\( \sigma_x= 375.854, \sigma_y= 353.364\)) respectively.

\subsubsection{Implementation Details}
Implemented in Keras and accelerated with Google Colaboratory Cloud P100 GPU, the proposed scheme is structured to input images with (256, 256, 3) resolution. Since there is class imbalance in the dataset, it was divided into 3 folds. Each fold included the 1082 normal class images. But from the opacity class of 3110 images, the k-th fold possessed the images within the range \(\floor*{3110/3}*(k-1)\) to \(\floor*{3110/3}*k\). Any remainder images are included in the final fold. Since there are multiple parallel branches, batch size is set to 2 with Adam(lr=1.5e-3) optimizer.



\subsubsection{Evaluation Metrics}
Important metrics such as Accuracy (Acc), Sensitivity (Sen), Specificity (Spe) and F1-score  are chosen to evaluate our methodology quantitatively.

\begin{equation}
\\Acc=\frac{TP+TN}{TP+FP+TN+FN}\
\end{equation}

\begin{equation}
\\Sen=\frac{TP}{TP+FN}\ 
\end{equation}

\begin{equation}
\\Spe=\frac{TN}{TN+FP}\ 
\end{equation} 

\begin{equation}
\\F1\;score=\frac{2*TP}{2*TP+FP+FN}\ 
\end{equation} 
 
where TP, FN, TN, and FP represent true positive, false negative, true negative and false positive, respectively. Furthermore, we introduce another metric to introduce comparison of the accuracy and parameter (in Millions) trade-off between various models, namely APT (accuracy parameter trade-off):
\begin{equation}
APT= 0.9\times Acc - 0.1\times \frac{Params}{7.3}
\end{equation}

The values `Acc' and `Params' are weighted (0.9 and 0.1) by us according to their relative significance for a nominal model. Also, `Params' is normalized with respect to our model parameters (7.3M).
\subsection{Performance of Proposed Framework}
The quantitative performance of the proposed method is analyzed and the comparison metric of it is graphed alongside a few other models in this section.

\subsubsection{Quantitative Performance Analysis}
Each fold of the dataset is trained on and the validation set is evaluated with the mentioned metrics detailed in Table~\ref{t2}. 

\begin{table}[htbp]
\caption{Performance Evaluation Table}
\label{t2}
\begin{center}
\begin{tabular}{ | c | c| c |c | c | } 
\hline
\textbf{Data Partition} & {\textbf{Accuracy}} & {\textbf{Sensitivity}} & {\textbf{Specificity}} &{\textbf{F1 score}}\\
\hline
Fold-1 & 0.9855 & 0.9922  & 0.9674 & 0.9903 \\ 
\hline
Fold-2 & 0.9788 & 0.9884 & 0.9535 &  0.986  \\ 
\hline
Fold-3 & 0.9692 & 0.9847 & 0.9303 & 0.9797 \\ 
\hline
\textbf{Mean} & \textbf{0.9778} & \textbf{0.9884} & \textbf{0.9504} & \textbf{0.9853} \\ 
\hline
\end{tabular}
\end{center}
\end{table}

\subsubsection{Qualitative Comparison Analysis}
\begin{figure}[!h]
  \includegraphics[width=\columnwidth]{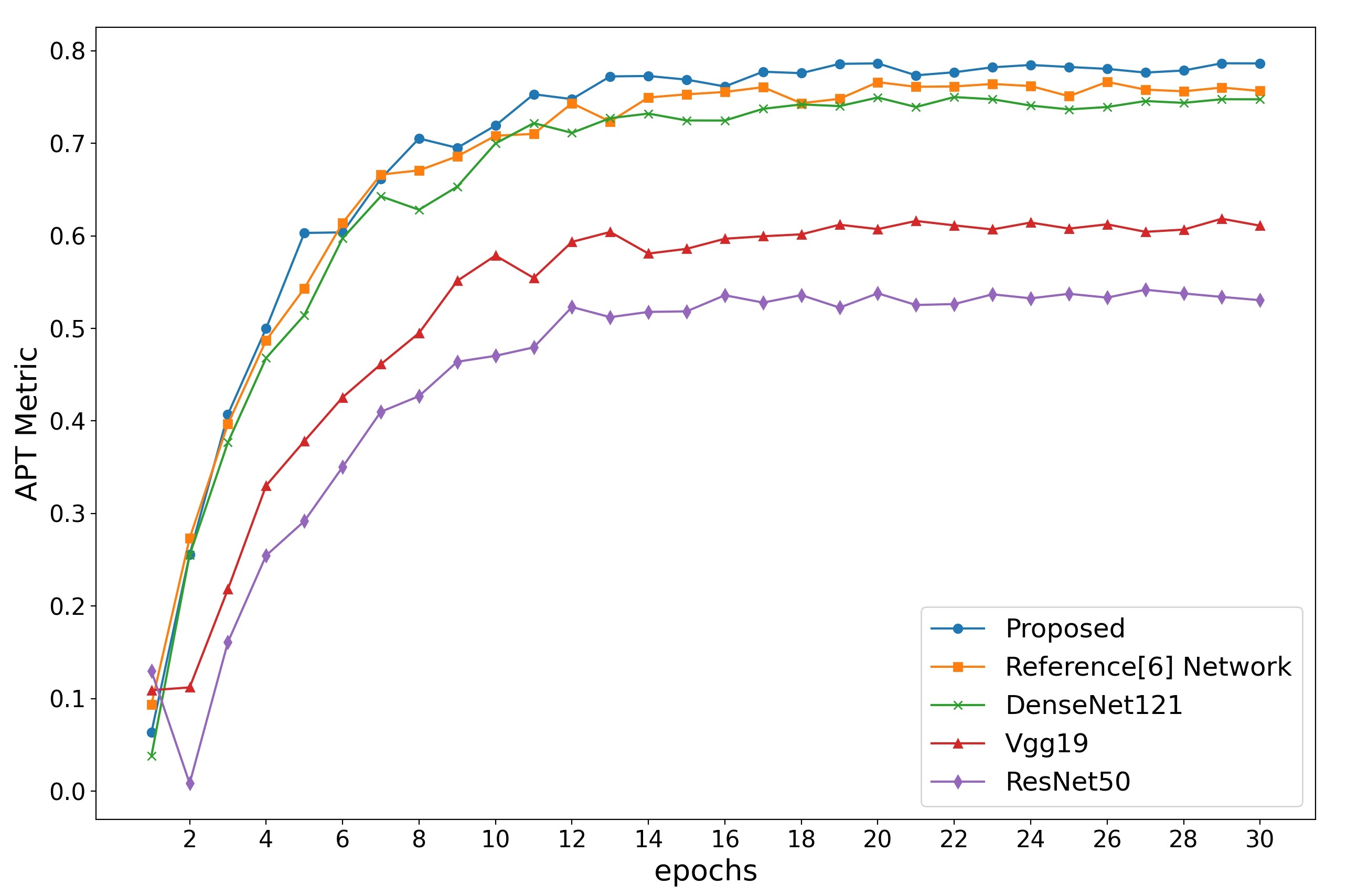}
  \caption{\textbf{APT Analysis for various models}}
  \label{fig3}
\end{figure}

To compare between relative models for pneumonia opacity detection, not only the accuracy but also the parameter count comes into consideration. The model presented by Sharma \textit{et al.}~\cite{sharma2020feature} has known accuracy and parameter count which are 0.9068 and 3.47M and therefore it was attempted to be reproduced for this experiment. Moreover, noted models such as VGG19 \cite{simonyan2014very}, ResNet50 \cite{he2016deep}, DenseNet121 \cite{huang2017densely} are also included for APT analysis. The representation of the qualitative analysis is portrayed in Fig.~\ref{fig3} for epochs up to 30 during which all the networks saturate.

\subsection{Ablation Studies}
In the proposed method, DF blocks are constructed by concatenating three separate branches with \(p1,\; p2,\; p3\). Here, the pathway bearing all `\(p1\)'s resembling a shallow VGG19 network, is the main forward stem and unavoidable. \textbf{Accuracy} values for including and excluding the other two paths are presented in Table~\ref{t3} for each fold of the data.
\par
Without the support of neither \(p2\) nor \(p3\), the model can not learn any distinguishing opacity attributes as seen in the table above. But if only \(p2\) pipelines are omitted, the pathways including \(p3\) contribute high level accuracy (an average of 0.9608) and significant feature accumulation as well as field of reception of image detail. Only \(p2\) alone is not enough for the network to learn the X-ray image because it possesses lower dimensional kernel (mean accuracy 0.2538). However, inclusion of both pathways increases the mean accuracy to 0.9778 as the \(p2\) layers provide smaller details to the model during concatenation, enhancing the performance compared to \(p3\) just being alone.

\begin{table}[h!]
\caption{Ablation Study of Accuracy for the Proposed Scheme}
\begin{center}
\label{t3}
\begin{tabular}{ | c | c| c |c | c | c | } 
\hline
\multicolumn{2}{|c|}{\textbf{p1 pathway always present}} & \multicolumn{4}{|c|}{\textbf{Accuracy Outcomes}}\\
\hline
\textbf{p2 pathways} & {\textbf{p3 pathways}} & {\textbf{Fold-1}} & {\textbf{Fold-2}} &{\textbf{Fold-3}} &{\textbf{Mean}}\\
\hline
\xmark & \xmark & 0.2577  & 0.253 & 0.2509 & 0.2538\\ 
\hline
\xmark & \cmark & 0.9672 & 0.9615 &  0.9537 & 0.9608 \\ 
\hline
\cmark & \xmark & 0.2577 & 0.2548 & 0.25 & 0.2542\\ 
\hline
\cmark & \cmark & \textbf{0.9855} & \textbf{0.9788} & \textbf{0.9692} & \textbf{0.9778}\\ 
 
\hline
\end{tabular}
\end{center}
\end{table}

\subsection{Computational Efficiency of the Network}
The size of the network as well as the iteration time per batch are effective indicators of the computational cost of any model. Parallel branches increase the iteration time; so similar models like ResNet50 and DenseNet121 is compared for this case with the proposed model. For a batch size of 2 in the same environment mentioned in the implementation details, the proposed network iterates each step within 280 ms whereas ResNet50 and DenseNet121 do it within 429 ms and 313 ms respectively, making our model faster. However, as far as the size of the network goes, the VGG19 model is estimated to be 85MB whereas our model parameters make it close to 32MB, making it lighter.

\section{Scopes for Future Works}
In future work, we aim to improve the classification accuracy of the model by fine tuning every parameter and hyper-parameter and also reducing layers if possible. Since mobile application bases solutions are very trendy this model could be quite handy. Model presented in this paper can be also extended to classify other diseases as CheXNet did with high accuracy. Overall performance of the model can be improved with the use of larger datasets with high resolution images.

\section{Conclusion}
This paper proposes a robust CNN based framework for detecting pneumonia that shows the use of a lightweight deep learning based framework which achieves high qualitative accuracy compared to the conventional methods. Simulation results show that the proposed Deep Learning based scheme boasts a mean accuracy of 97.78\%, sensitivity 98.84\%,	specificity 95.04\%. The obtained sensitivity and specificity values imply that on the given validation set there are relatively higher false positives (Fold-1: 13, Fold-2: 20, Fold-3: 9) rather than false negatives (Fold-1: 9, Fold-2: 12, Fold-3: 6) which is practically less fatal than the other way around. This kind of lightweight network can motivate mobile applications of similar detection systems. In medical applications with such systems involved, patients can know their diagnosis without even going to hospitals.

\bibliographystyle{IEEEtran}
\bibliography{ref}

\begin{thebibliography}{10}
\providecommand{\url}[1]{#1}
\csname url@samestyle\endcsname
\providecommand{\newblock}{\relax}
\providecommand{\bibinfo}[2]{#2}
\providecommand{\BIBentrySTDinterwordspacing}{\spaceskip=0pt\relax}
\providecommand{\BIBentryALTinterwordstretchfactor}{4}
\providecommand{\BIBentryALTinterwordspacing}{\spaceskip=\fontdimen2\font plus
\BIBentryALTinterwordstretchfactor\fontdimen3\font minus
  \fontdimen4\font\relax}
\providecommand{\BIBforeignlanguage}[2]{{%
\expandafter\ifx\csname l@#1\endcsname\relax
\typeout{** WARNING: IEEEtran.bst: No hyphenation pattern has been}%
\typeout{** loaded for the language `#1'. Using the pattern for}%
\typeout{** the default language instead.}%
\else
\language=\csname l@#1\endcsname
\fi
#2}}
\providecommand{\BIBdecl}{\relax}
\BIBdecl

\bibitem{bates1992microbial}
J.~H. Bates, G.~D. Campbell, A.~L. Barton, G.~A. McCracken, P.~N. Morgan, E.~B.
  Moses, and C.~M. Davis, ``Microbial etiology of acute pneumonia in
  hospitalized patients,'' \emph{Chest}, vol. 101, no.~4, pp. 1005--1012, 1992.

\bibitem{info}
``Top 20 pneumonia facts—2019,'' 2019, [Online]. Available:
  \url{https://www.thoracic.org/patients/patient-resources/resources/top-pneumonia-facts.pdf}.

\bibitem{jaiswal2019}
A.~K. Jaiswal, P.~Tiwari, S.~Kumar, D.~Gupta, A.~Khanna, and J.~J. Rodrigues,
  ``Identifying pneumonia in chest x-rays: A deep learning approach,''
  \emph{Measurement}, vol. 145, pp. 511--518, 2019.

\bibitem{chapman2001comparison}
W.~W. Chapman, M.~Fizman, B.~E. Chapman, and P.~J. Haug, ``A comparison of
  classification algorithms to automatically identify chest x-ray reports that
  support pneumonia,'' \emph{Journal of biomedical informatics}, vol.~34,
  no.~1, pp. 4--14, 2001.

\bibitem{rajpurkar2017chexnet}
P.~Rajpurkar, J.~Irvin, K.~Zhu, B.~Yang, H.~Mehta, T.~Duan, D.~Ding, A.~Bagul,
  C.~Langlotz, K.~Shpanskaya \emph{et~al.}, ``Chexnet: Radiologist-level
  pneumonia detection on chest x-rays with deep learning,'' \emph{arXiv
  preprint arXiv:1711.05225}, 2017.

\bibitem{sharma2020feature}
H.~Sharma, J.~S. Jain, P.~Bansal, and S.~Gupta, ``Feature extraction and
  classification of chest x-ray images using cnn to detect pneumonia,'' in
  \emph{2020 10th International Conference on Cloud Computing, Data Science \&
  Engineering (Confluence)}.\hskip 1em plus 0.5em minus 0.4em\relax IEEE, 2020,
  pp. 227--231.

\bibitem{chandra2020pneumonia}
T.~B. Chandra and K.~Verma, ``Pneumonia detection on chest x-ray using machine
  learning paradigm,'' in \emph{Proceedings of 3rd International Conference on
  Computer Vision and Image Processing}.\hskip 1em plus 0.5em minus 0.4em\relax
  Springer, 2020, pp. 21--33.

\bibitem{qin2018computer}
C.~Qin, D.~Yao, Y.~Shi, and Z.~Song, ``Computer-aided detection in chest
  radiography based on artificial intelligence: a survey,'' \emph{Biomedical
  engineering online}, vol.~17, no.~1, p. 113, 2018.

\bibitem{li2020accuracy}
Y.~Li, Z.~Zhang, C.~Dai, Q.~Dong, and S.~Badrigilan, ``Accuracy of deep
  learning for automated detection of pneumonia using chest x-ray images: A
  systematic review and meta-analysis,'' \emph{Computers in Biology and
  Medicine}, p. 103898, 2020.

\bibitem{xu2018quantization}
X.~Xu, Q.~Lu, L.~Yang, S.~Hu, D.~Chen, Y.~Hu, and Y.~Shi, ``Quantization of
  fully convolutional networks for accurate biomedical image segmentation,'' in
  \emph{Proceedings of the IEEE conference on computer vision and pattern
  recognition}, 2018, pp. 8300--8308.

\bibitem{acharya2020deep}
J.~Acharya and A.~Basu, ``Deep neural network for respiratory sound
  classification in wearable devices enabled by patient specific model
  tuning,'' \emph{IEEE transactions on biomedical circuits and systems},
  vol.~14, no.~3, pp. 535--544, 2020.

\bibitem{cherupally2019ecg}
S.~K. Cherupally, G.~Srivastava, S.~Yin, D.~Kadetotad, C.~Bae, S.~J. Kim, and
  J.-s. Seo, ``Ecg authentication neural network hardware design with
  collective optimization of low precision and structured compression,'' in
  \emph{2019 IEEE International Symposium on Circuits and Systems
  (ISCAS)}.\hskip 1em plus 0.5em minus 0.4em\relax IEEE, 2019, pp. 1--5.

\bibitem{shuvo2020}
S.~B. Shuvo, S.~N. Ali, S.~I. Swapnil, T.~Hasan, and M.~I.~H. Bhuiyan, ``A
  lightweight cnn model for detecting respiratory diseases from lung
  auscultation sounds using emd-cwt-based hybrid scalogram,'' \emph{arXiv
  preprint arXiv:2009.04402}, 2020.

\bibitem{xiao2020heart}
B.~Xiao, Y.~Xu, X.~Bi, J.~Zhang, and X.~Ma, ``Heart sounds classification using
  a novel 1-d convolutional neural network with extremely low parameter
  consumption,'' \emph{Neurocomputing}, vol. 392, pp. 153--159, 2020.

\bibitem{simonyan2014very}
K.~Simonyan and A.~Zisserman, ``Very deep convolutional networks for
  large-scale image recognition,'' \emph{arXiv preprint arXiv:1409.1556}, 2014.

\bibitem{he2016deep}
K.~He, X.~Zhang, S.~Ren, and J.~Sun, ``Deep residual learning for image
  recognition,'' in \emph{Proceedings of the IEEE conference on computer vision
  and pattern recognition}, 2016, pp. 770--778.

\bibitem{d1}
``{Chest X-Ray Images (Pneumonia)},'' 2020, [Online]. Available:
  \url{https://www.kaggle.com/pcbreviglieri/pneumonia-xray-images}.

\bibitem{huang2017densely}
G.~Huang, Z.~Liu, L.~Van Der~Maaten, and K.~Q. Weinberger, ``Densely connected
  convolutional networks,'' in \emph{Proceedings of the IEEE conference on
  computer vision and pattern recognition}, 2017, pp. 4700--4708.

\end{thebibliography}

\end{document}